\documentclass[aps,amsmath,amssymb,twocolumn,pra,10pt]{revtex4-2}

\usepackage{color}
\usepackage{physics}
\usepackage{amsthm}
\usepackage{bbm}
\usepackage{tikz}
\usepackage{xcolor}
\usepackage{soul}
\usepackage{inputenc}
\usepackage{babel}

\usepackage[compatibility=false]{caption}
\usepackage{graphicx}

\usepackage{dcolumn}
\usepackage{bm}

\usepackage{caption}
\usepackage{url}
\usepackage{subcaption}
\usepackage[justification=centering]{caption}
\usepackage{hyperref}

\begin{document}

\title{Fluorescence spectrum of a hybrid three-level quantum dot-nanoparticle system}

\author{Aryan Iliat}

\affiliation{School of Physics and Applied Physics, Southern Illinois University, Carbondale, IL, 62903, USA}

\date{\today}

\begin{abstract}
Quantum optics provides a fundamental framework for understanding the interaction between light and matter at the quantum level.  Recently, it has been shown that under incoherent pumping, the
resonance fluorescence spectrum dramatically changes. Engineering the resonance
fluorescence spectrum paves the way towards solid-state-based single-photon sources.
In this paper, we start by reviewing and reproducing some of the results concerning the resonance
fluorescence spectrum, single-photon sources, dressed-state lasers, and luminescence spectrum of a quantum dot in a microcavity. Photon correlations in quantum optical systems and spectral properties
of radiation emitted by atomic and semiconductor systems interacting with external fields are investigated. The well-known Mollow triplet structure of the emission spectrum is
discussed, together with the role of dressed states in explaining the origin of the three spectral
peaks. 
Furthermore, the luminescence spectra of quantum emitters coupled to microcavities are reviewed. 
The numerical results presented here contribute to the theoretical understanding of resonance fluorescence, photon correlations, and engineered emission in quantum optical systems. These studies highlight the rich physical properties arising from light–matter interaction at the quantum level and demonstrate their relevance for emerging quantum technologies.

\end{abstract}

\maketitle

\section{\label{sec:background}Introduction}
One of the most important phenomena in quantum physics is the interaction of light with matter. 
The photon emerged as a theoretical concept for describing the fundamental properties of the electromagnetic field \cite{10.1119/1.1971542}. Among these properties are the relationship between the energy and frequency of light, the thermal equilibrium between light and matter, and the photoelectric effect \cite{https://doi.org/10.1002/andp.19053220607}. With the advances made in the generation, propagation, and detection of photons, quantum systems are now investigated at the single-photon level. Therefore, there is a strong need to extend and refine the theory of photon detection \cite{vogel2006quantum}. 
For example, obtaining photon correlations that include temporal and frequency information has nowadays become common in laboratories \cite{Ma:11, Kuo:12}.
On a microscopic scale, the light-matter interaction  can be described in two ways: semiclassically or quantum mechanically \cite{Scully_Zubairy_1997}.
In both approaches, matter is treated quantum mechanically using the Schrödinger equation. In the simplest model, matter is considered a two-level system (qubits). However, multilevel systems can also be used for this purpose (qudits). Light, on the other hand, can be described either classically or quantum mechanically. For example, when describing a strong laser field, the classical description is often sufficient \cite{PhysRev.181.618}.
However, certain phenomena cannot be explained using the semiclassical theory of light. For instance, to explain spontaneous emission from atomic levels or the Casimir effect \cite{Lamoreaux_2005, RevModPhys.81.1827}, it is necessary to employ the quantum theory of light.

The resonance fluorescence spectrum of a two-level atom interacting with classical light was first studied by Mollow \cite{PhysRevB.79.235326}. The Mollow spectrum of coherent light scattered from a two-level atom has three
peaks at $\omega_0$ and $\omega_0 \pm \Omega$, where $\omega_0$ is the atom transition frequency and $\Omega$ is the
Rabi frequency.  Such systems are particularly important for the realization of solid–state single–photon sources and their applications in quantum communication and quantum information processing \cite{PhysRevB.101.081401, https://doi.org/10.1002/adfm.202315936}. 
In this context, a quantum dot is of great interest to many researchers as a two-level system. A key problem is the interaction of a quantum dot inside a microcavity with quantized light. In Refs.~\cite{PhysRevB.79.235326, PhysRevB.79.235325, PhysRevB.73.115343}, it has been shown that incoherent pumping strongly modifies the fluorescence spectrum. 

In this paper, we start by reviewing the spectrum of an atom coupled to a cavity in \ref{sec:flointspec}. 
Section \ref{sec:photon} reviews the statistics of photons. It explains the separation of a classical and a quantum case. Here, it can be seen that when a single-photon source can be achieved, and discusses the equivalence of first- and
second-order correlation functions for open quantum systems. Section \ref{sec:filterspec} shows a more rigorous way to get single photons from a system by analyzing the correlations between photons. 
Three-level state systems were also mentioned as an alternative to the widely used two-level systems.  Section \ref{sec:dressedlaser} includes a review of the two-level system for producing dressed-state lasers \cite{PhysRevA.44.7717}, where the atom-cavity Hamiltonian is written in the rotating-wave approximation and laser dynamics are obtained via the Lindblad equations. Finally, in section \ref{sec:calculation}, an example of these systems for a three-level case was numerically calculated, and the results show the correlation functions and their spectra. 
%----------------------------------------------------------------------------------------------
\section{Fluorescence Intensity Spectrum}\label{sec:flointspec}
In 1969, Mollow obtained the scattering spectrum of light from a stationary two-level atomic system. In this study, the classical incident light frequency was assumed to be approximately equal to the atomic transition frequency \cite{PhysRev.188.1969}.   
The interaction of the atom with the field is given by
\begin{eqnarray}\label{agha1}
    \mathbf{H}_I(t)=-\mathbf{d}\cdot\mathbf{E}(t),
\end{eqnarray}
where  
$\mathbf{d}$  
is the electric dipole moment of the atom and  
$\mathbf{E}(t)$  
is the electric field at the location of the atom at  time  
$t$ \cite{PhysRev.188.1969}.  
Writing Eq.~\eqref{agha1} in the rotating frame, the atom-field interaction Hamiltonian becomes
\begin{eqnarray}\label{agha2}
    \text{H}_I=\Omega_L(\sigma+\sigma^\dagger),
\end{eqnarray}
where $\sigma=\ket{0}\bra{1}$ is the atomic lowering operator, and the Rabi frequency $\Omega_L$ is proportional to the magnitude of the electric field \cite{PhysRevLett.105.233601}.  Here,  
$\ket{0}$  
is the ground state of the two-level system with zero energy, and  
$\ket{1}$  
is the excited state with energy  
$\hbar \omega_0$. 
The Lindblad equation
\begin{eqnarray*}
\partial_t \rho = i[\rho,\text{H}_I] + \frac{\gamma_\sigma}{2} \mathcal{L}_\sigma(\rho)
\end{eqnarray*}
is used to describe the effect of the environment (vacuum electromagnetic field modes) on the system, where $\gamma_\sigma$ is the decay rate \cite{PhysRevLett.105.233601} and
\begin{eqnarray*}
\mathcal{L}_\sigma(\rho) = 2\sigma \rho \sigma^\dagger - \sigma^\dagger \sigma \rho - \rho \sigma^\dagger \sigma.
\end{eqnarray*}  

In the fluorescence spectrum, three peaks can be observed where 
the splitting of the energy levels is due to the dynamical Stark effect \cite{PhysRev.100.703, duncan2014troubleorbitsstarkeffect}. 
The central peak of the fluorescence spectrum occurs at the same frequency as the transition frequency between the atomic levels. 
The side peaks of the spectrum arise from transitions between the dressed states (atomic states shifted due to the presence of the external field) \cite{PhysRevA.84.043816, PAlsing_1991, PhysRevResearch.6.043247}.
It can be shown that the height of the central peak is three times larger than that of the side peaks, and its width is 1.5 times smaller than the width of the side peaks \cite{PhysRev.188.1969}.

Recently, the effect of incoherent pumping of atomic levels on the Mollow spectrum has been investigated \cite{PhysRevLett.105.233601, PhysRevA.98.013820}.  
Let's assume the two-level system (a quantum dot, for example) is located inside a microcavity \cite{lambropoulos2007fundamentals}. Due to the presence of the cavity, it is sufficient to consider only a single quantized mode of the electromagnetic field. This mode is described by the Hamiltonian  
$\hbar \omega a^\dagger a$.  

The combined light–two-level system interaction Hamiltonian is
\begin{eqnarray*}
H_I = g(a^\dagger \sigma + a \sigma^\dagger),
\end{eqnarray*}
where $g$ is the coupling constant between the quantum dot and the cavity mode. The system evolution equation is
\begin{eqnarray}\label{agha3}
\partial_t \rho = i[\rho,H] + \frac{\gamma_a}{2}\mathcal{L}_a(\rho) + \frac{\gamma_\sigma}{2}\mathcal{L}_\sigma(\rho) + \frac{P_\sigma}{2}\mathcal{L}_\sigma^\dagger(\rho).
\end{eqnarray}
The incoherent pumping introduces the term 
$\frac{P_\sigma}{2}\mathcal{L}_\sigma^\dagger(\rho)$  
in the evolution equation. Ultimately, this evolution leads to the incoherent fluorescence spectrum\cite{PhysRevLett.105.233601}. 

\section{Photon Statistics}\label{sec:photon}
One of the results of quantum optics is that it shows there are three different types of photon statistics for a single light beam \cite{z3cr-l7pw}:
(a) Poissonian,  
(b) sub-Poissonian \cite{Mandel:79, PhysRevLett.51.384, Dmitriev_2025},  and
(c) super-Poissonian \cite{fox2006quantum, Kovalenko:23, Alzar_2003, PhysRevLett.110.117401}.  
Poissonian and super-Poissonian photon statistics can be classically expected. The existence of sub-Poissonian statistics for a light beam is one of the remarkable predictions of the quantum theory of light and has been experimentally verified \cite{fox2006quantum}.  

Consider a light beam, and let $n$ be the random variable representing the number of photons in a given time interval (this is the propagation length). The probability distribution function of $n$ for Poissonian light is \cite{alishah_probability, PhysRevA.94.053844, Zeller_2020}
\begin{equation*}
\mathcal{P}(n) = \frac{\bar{n}^n}{n!} \exp(-\bar{n}), \quad n = 0, 1, 2, \dots .
\end{equation*}
For example, perfectly coherent light with constant intensity, whose electric field 
\begin{equation*}
\mathcal{E}(x,t) = \mathcal{E}_0 \sin(kx - \omega t + \Phi),
\end{equation*}
follows Poissonian statistics. However, 
thermal light (electromagnetic radiation of a hot body, also called blackbody radiation) exhibits super-Poissonian statistics.  
For a probability distribution $P(n)$, the mean value $\bar{n}$ and standard deviation $\Delta n$ are defined as
\begin{eqnarray*}
\bar{n} = \sum n P(n), \,\text{and}\quad
(\Delta n)^2 = \sum (n-\bar{n})^2 P(n).
\end{eqnarray*}
For a Poissonian distribution, $\Delta n = \bar{n}$, for super-Poissonian $\Delta n > \bar{n}$, and $\Delta n < \bar{n}$ for sub-Poissonian distribution.

Another method for classifying light beams is based on the second-order correlation function $g^{(2)}(t^\prime)$ \cite{fox2006quantum, 1995ocqo.book.M, Grunwald_2019, PhysRev.130.2529, 7ckm-tm3r}, which is defined as
\begin{equation}\label{agha5}
g^{(2)}(t^\prime) \!=\! \frac{\langle \mathcal{E}^*(t)\mathcal{E}^*(t\!+\!t^\prime)\mathcal{E}(t\!+\!t^\prime)\mathcal{E}(t) \rangle}{\langle \mathcal{E}^*(t)\mathcal{E}(t) \rangle \langle \mathcal{E}^*(t\!+\!t^\prime)\mathcal{E}(t\!+\!t^\prime) \rangle} \!=\! \frac{\langle I(t) I(t\!+\!t^\prime) \rangle}{\langle I(t) \rangle \langle I(t\!+\!t^\prime) \rangle}.
\end{equation}
Here $\mathcal{E}(t)$ and $I(t)$ are the electric field and intensity of the light beam at time $t$. This correlation function is measured using a Hanbury Brown and Twiss interferometer \cite{1956Natur177_27B}.  These tools allow the investigation of photon statistics that contain both temporal and frequency
information, which has become experimentally accessible with modern detection techniques.
Based on the value of $g^{(2)}(t^\prime)$, light can be classified as:  
(a) bunched light, $g^{(2)}(0) > 1$,  
(b) coherent light, $g^{(2)}(0) = 1$, and 
(c) anti-bunched light, $g^{(2)}(0) < 1$.  

Bunched light emits photons in clusters: if a photon is detected at $t=0$, it is more likely that another photon will be detected shortly after than at later times. Anti-bunched light exhibits nearly regular spacing between photons.   Coherent light, on the other hand, has random spacing between photons.

Anti-bunched single-photon light is particularly important for quantum technologies, quantum communication \cite{Gisin_2007}, and quantum cryptography \cite{PhysRevA.54.97, Bennett:1992xkm}.
Many single-photon sources are based on parametric down-conversion \cite{vogel2006quantum, PhysRevLett.118.030503}.

Another method to generate single photons is filtering the Mollow spectrum, which will be explained later.
We want to pay particular attention to the topics of resonance fluorescence and photon statistics, both of which reveal the quantum nature of light.
Here, we look at the spectral and statistical properties of photons
emitted from a cavity coupled to two quantum emitters. 

The dynamics of the system are described using the Lindblad master equation

\begin{equation}
\frac{d\rho}{dt} = -i[H,\rho] + \sum_k \frac{\gamma_k}{2}
\left(2L_k \rho L_k^\dagger - L_k^\dagger L_k \rho - \rho L_k^\dagger L_k\right).
\end{equation}
The Hamiltonian of the cavity–emitter system is
\begin{equation}
H = \omega_a a^\dagger a
+ \sum_{i=1}^{2} \omega_{\sigma_i} \sigma_i^\dagger \sigma_i
+ \sum_{i=1}^{2} g_i (a^\dagger \sigma_i + a \sigma_i^\dagger),
\end{equation}
where $\omega_a$ is the electromagnetic mode, $\omega_{\sigma_i}$ is the frequency of the physical system used in the experiment, and $g_i$ is the coupling constant between them.

\section{Filtering the Fluorescence Resonance Spectrum}\label{sec:filterspec}
As mentioned before, one way to analyze correlations between photons is by filtering the fluorescence resonance spectrum
\cite{PhysRevA.47.510, Laussy2017QuantumOA},
which allows the production of single-photon emission or cascaded photon emission
\cite{2012NaPho.6.238U}. Recall that the Mollow spectrum has three separate peaks.
The side peaks of the quantum dot fluorescence spectrum are separated and directed into interferometric devices placed before the detectors
\cite{Valle_2013}. 
 
The two-photon correlation is obtained using the relation
\begin{equation}\label{agha6}
g^{(2)}_{\Gamma_1\Gamma_2}(\omega_1,t^\prime_1;\omega_2,t^\prime_2)
=
\frac{S^{(2)}_{\Gamma_1\Gamma_2}(\omega_1,t^\prime_1;\omega_2,t^\prime_2)}
{S^{(1)}_{\Gamma_1}(\omega_1,t^\prime_1)S^{(1)}_{\Gamma_2}(\omega_2,t^\prime_2)}
\end{equation}
\cite{PhysRev.130.2529, PhysRev.130.2529},
where \cite{Eberly:77}
\begin{widetext}
\begin{eqnarray*}
S^{(1)}_{\Gamma_1}(\omega_1,t^\prime_1)&=&2\Gamma_1\int\int_{-\infty}^{t^\prime_1} dt_1 dt_2\, e^{-(\Gamma_1-i\omega_1)(t^\prime_1-t_1)} e^{-(\Gamma_1+i\omega_1)(t^\prime_1-t_2)} \langle a^\dagger(t_1) a(t_2) \rangle,\\
S^{(1)}_{\Gamma_2}(\omega_2,t^\prime_2)&=&2\Gamma_2\int\int_{-\infty}^{t^\prime_2} dt_1 dt_2\, e^{-(\Gamma_2+i\omega_2)(t^\prime_2-t_1)} e^{-(\Gamma_2-i\omega_2)(t^\prime_2-t_2)} \langle a(t_1) a^\dagger(t_2) \rangle,\\
S^{(2)}_{\Gamma_1,t^\prime_1,\Gamma_2,t^\prime_2}(\omega_1,t^\prime_1; \omega_2,t^\prime_2)
&=&4 \Gamma_1 \Gamma_2 \int\int_{-\infty}^{t^\prime_1} dt_1 dt_2\, e^{-\Gamma_1(t^\prime_1-t_1)} e^{-\Gamma_1(t^\prime_1-t_2)} \\
& \times& \int\int_{-\infty}^{t^\prime_2} dt_3 dt_4\, e^{-\Gamma_2(t^\prime_2-t_3)} e^{-\Gamma_2(t^\prime_2-t_4)} e^{i\omega_1(t_2-t_1)} e^{i\omega_2(t_4-t_3)} 
 \langle a^\dagger(t_1) a(t_2) a(t_3) a^\dagger(t_4) \rangle.
\end{eqnarray*}
\end{widetext}
In the above equations, $\Gamma_1$ and $\Gamma_2$ correspond to the frequency bandwidths of the detectors, $t^\prime_1$ and $t^\prime_2$ denote the times when the detectors detect the respective photons. That is, whenever a photon is emitted from the first source (with $t^\prime_1=0$) it reaches the first detector at time $t^\prime_1$.  This spectrum, in the physical sense, is defined in the steady-state regime and, in the limit $\Gamma_1 \to 0$, reduces to the Wiener-Khinchin theorem \cite{wiener1930generalized, Khintchine1934KorrelationstheorieDS, Jin:18}.

The generalization of this result to the case of two-photon detection was inspired by  \cite{PhysRevLett.45.617}  on resonance fluorescence in the Mollow triplet regime  \cite{cohentannoudji1979AtomsIS, reynaud1983fluorescence, dalibard1983correlation}.
The extension of photon detection using the concept introduced by Eberly and Wódkiewicz—namely, considering two detectors with bandwidths $\Gamma_1$ and $\Gamma_2$—was carried out by Knoll \cite{LKnoll_1984}  and by Arnoldus and Nienhuis \cite{HFArnoldus_1984}.

By generalizing equation \eqref{agha6}, the Nth-order correlations between filtered photons can also be calculated. In practice, however, the calculations become highly complex even in the case of the single-mode systems \cite{PhysRev.130.2529, PhysRevA.42.503, NEELEN1993289}. Therefore, a new method introduced in \cite{PhysRevLett.109.183601} has recently attracted significant attention. In this method, $N$ sensors are weakly coupled to an open quantum system, as shown in  \cite{PhysRevLett.109.183601}.
Here, the $i$-th sensor is a two-level system with transition frequency $\omega_i$ and linewidth $\Gamma_i$. It is proven that \cite{PhysRevLett.109.183601}
\begin{eqnarray}
g^{(N)}_{\Gamma_1\cdots\Gamma_N}(\omega_1,t^\prime_1;\cdots;\omega_N,t^\prime_N)
\!=\!\!\!\!\!\!
\lim_{\varepsilon_1,\cdots,\varepsilon_N\to0} \!
\frac{\langle n_1(t^\prime_1)\cdots n_N(t^\prime_N)\rangle}{\langle n_1(t^\prime_1)\rangle \cdots \langle n_N(t^\prime_N)\rangle}.\notag
\end{eqnarray}
In particular, when $N=2$,
\begin{equation}
g^{(2)}_{\Gamma_1\Gamma_2}(\omega_1,t^\prime_1;\omega_2,t^\prime_2)
=
\lim_{\varepsilon_1,\varepsilon_2\to0} 
\frac{\langle n_1(t^\prime_1) n_2(t^\prime_2) \rangle}{\langle n_1(t^\prime_1)\rangle \langle n_2(t^\prime_2)\rangle},
\end{equation}
where
\begin{eqnarray*}
\langle n_i \rangle \!=\! \frac{2\pi\varepsilon_i^2}{\Gamma_i}  S_{\Gamma_i}^{(1)}(\omega_i),\quad
\langle n_1 n_2 \rangle \!=\! \frac{(2\pi\varepsilon_1 \varepsilon_2)^2}{\Gamma_1 \Gamma_2} S^{(2)}_{\Gamma_1;\Gamma_2}(\omega_1, \omega_2),
\end{eqnarray*}
for simultaneous photon detection, and
\begin{equation*}
\langle n_1(0) n_2(t^\prime) \rangle = \frac{\varepsilon_1^2 \varepsilon_2^2}{\Gamma_1 \Gamma_2} (2\pi)^2 S^{(2)}_{\Gamma_1 \Gamma_2}(\omega_1;\omega_2, t^\prime),
\end{equation*}
for photon detection with a time delay $t^\prime$. If the correlation function $g^{(2)}(t^\prime) < 1$, anti-bunched light is observed \cite{2026arXiv260210882T, Elmalem:25, PhysRevLett.93.096801, Boddeda_2019}, which means that this setup can be used as a single-photon source.

\begin{figure}[ht]
\includegraphics[width=0.495\textwidth]{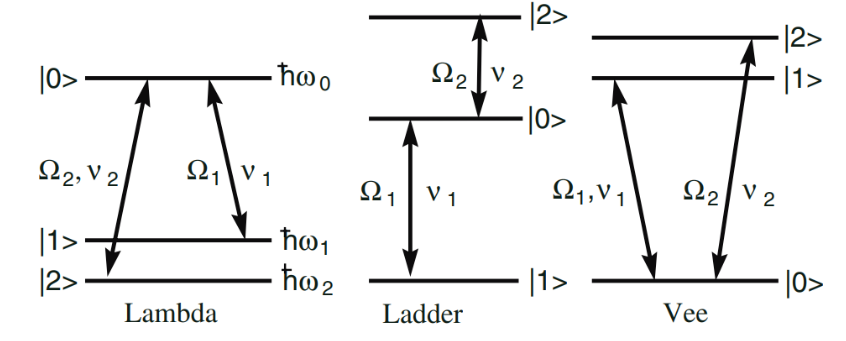}
	\caption{
	Various three-level systems under the action of two lasers with frequencies 
$\nu_1$ and $\nu_2$ are considered. The transition between the levels 
$\ket{1}$ and $\ket{2}$ is assumed to be forbidden. 
$\Omega_k$ is the Rabi frequency of the transition between the levels 
$\ket{0}$ and $\ket{k}$. The energy of each level is 
$\hbar \omega_k$ \cite{ficek2005quantum}.
	}
    \label{fig3ss}
\end{figure}

 Now let's explore three-level systems. These systems exhibit richer physics than the two-level system \cite{Yale_2016, PhysRevA.107.012424, Helmrich_2016, Gavryusev_2016, 10.1063/1.4936206, iliat2025physically, Zanner_2022, PhysRevApplied.11.014053, PhysRevB.87.214515, PhysRevLett.125.233605}. For example, Electromagnetically Induced Transparency (EIT) is notable \cite{1997PhT50g36H, RevModPhys.77.633, Finkelstein_2023, PhysRevLett.66.2593, Ma_2017, Wang_2020}.
The three-level fluorescence resonance spectrum is significant. Suppose the connection of the excited states $\ket{1}$ and $\ket{2}$ with the ground state $\ket{0}$ is via an electric dipole transition \cite{ficek2005quantum}. The cosine of the angle between the two dipole moments controls quantum interference between the two transition paths. Maximum quantum interference occurs if the dipoles are parallel, and minimum interference occurs if they are perpendicular. In \cite{alma991019425319706535}, a three-level system driven by non-monochromatic light with limited frequency bandwidth (not single-frequency) was studied.  
Depending on the type of a three-level system, its Hamiltonian can be slightly different.
The different types of three-level systems were shown in FIG. \ref{fig3ss} \cite{ficek2005quantum}.
In this paper, the Vee-type three-level system is studied.
\section{Dressed-state lasers}\label{sec:dressedlaser}
In this section, we review the theory of dressed-state lasers to better understand the atom-cavity interaction.
In recent years, considerable attention has been paid to optical instabilities that lead to the generation of various types of radiation. It has been shown that an increase in the population inversion between the sublevels of the dressed states makes it possible to realize a dressed-state laser \cite{PhysRevA.44.7717, Janjusevic:90, duchateau2017theoreticalderivationlaserdressedatomic, 2018NaPho.12.4D, 2014LaPhy.24g4006C, 2013PhRvA.88f3834C, 1992PhRvA.46.2877Z, 1992PhRvA.45.2057Z, 1991PhRvA.44.7732Z, 1991PhRvA.44.7746Z} which was proposed by Mollow. This type of laser operates through one-, two-, or multi-photon stimulated emission.  It was predicted that a strongly driven ensemble of homogeneous two-level atoms exhibits not only optical absorption but also optical gain. Examination of conventional laser systems shows that gain effects are enhanced by using an optical cavity. When the cavity contains a sufficient number of atoms, the gain can overcome losses, and a steady-state dressed-state laser, in which the two-level atoms form the gain medium, may occur.  

A two-level atom subjected to a strong laser field with frequency 
$\omega_L$ 
becomes dressed. If the frequency 
$\omega_L$ 
(the driving frequency) is detuned from the atomic transition frequency 
$\omega_a$ 
by 
$\Delta_1 = \omega_a - \omega_L$, 
and $\Omega$ is the Rabi frequency, the atom-field states form a ladder of doublets separated by the frequency 
$\omega_L$, 
with a generalized Rabi splitting of 
$\Omega^\prime = (\Delta_1^2 + \Omega^2)^{1/2}$. 
This type of laser has been observed with atomic vapor cells 
\cite{PhysRevLett.60.1126} 
and atomic beam experiments 
\cite{PhysRevA.41.1576}.
\subsection{Atom-field interaction in the dressed state basis}
We assume that there are 
$N$ 
two-level atoms inside a cavity, which are driven by an external field with frequency 
$\omega_{L}$. 
We aim to derive the dynamical equations for the macroscopic polarization, population inversion, and the cavity field amplitude. In these calculations, we employ the rotating wave approximation. The Hamiltonian of the system is given by:
\begin{eqnarray}\label{fa1}
H &=& \frac{1}{2} \sum_{\mu}^{N} \big[ \Delta_1 \sigma_{3\mu} + \Omega (\sigma_\mu + \sigma_\mu^\dagger) + g_\mu \sigma_\mu^{\dagger} a + g_\mu^\ast a^\dagger \sigma_\mu \big]\notag\\
&+& \Delta_2 a^\dagger a,
\end{eqnarray}
where 
$\sigma_{1\mu}$, 
$\sigma_{2\mu}$, and 
$\sigma_{3\mu}$ 
are the Pauli matrices ($X$, $Y$, and $Z$), and 
$\mathit{a}$ and 
$\mathit{a^\dagger}$ 
are the photon-cavity annihilation and creation operators. $\Omega$ is the Rabi frequency.  In order to eliminate the time dependence of the field at frequency $\omega_L$, we use the rotating wave approximation. Here $\omega_a$ is the atomic resonance frequency and $\omega_c$ is the cavity resonance frequency. We define
$\Delta_1=\omega_a-\omega_L$, $\Delta_2=\omega_c-\omega_L$, and
$g_\mu=g \mathit{e}^{i\phi_\mu}$.
The second term in Eq.~\eqref{fa1} corresponds to the driving of the atoms, while the third and fourth terms describe the coupling of atom $\mu$ to the cavity mode. 
The atomic raising and lowering operators, 
$\sigma_\mu^\dagger$ 
and 
$\sigma_\mu$, 
are defined as:
\begin{equation}\label{teh}
\sigma_\mu^\dagger = \frac{1}{2}(\sigma_{1\mu} + i \sigma_{2\mu}), \,\,\text{and} \quad
\sigma_\mu = \frac{1}{2}(\sigma_{1\mu} - i \sigma_{2\mu}).
\end{equation}
    
    The time evolution of the density matrix of the system, $\rho$, is given by the Liouville-von Neumann equation \cite{Neumann1927}:
\begin{equation}\label{ff}
\dot{\rho}=-\mathit{i}[H,\rho]+{L_A}\rho+{L_F}\rho
\end{equation}
where $\rho$ is the density matrix of the system and $H$ is its Hamiltonian.

In Eq.~\eqref{ff}, the effect of cavity damping is represented by $L_F \rho$, while the effect of spontaneous emission is described by $L_A \rho$:
\begin{eqnarray}
L_F \rho &=& 2\Gamma\left[\mathit{a}\rho \mathit{a}^\dagger-\frac12 a^\dagger a \rho-\frac12\rho \mathit{a}^\dagger \mathit{a}\right] \label{ff2}\\
L_A \rho &= &2\gamma\sum_{\mu}\left[\sigma_\mu \rho\sigma_\mu^\dagger-\frac12\sigma_\mu^\dagger\sigma_\mu\rho-\frac12\rho\sigma_\mu^\dagger\sigma_\mu\right] \label{ff3}
\end{eqnarray}

In Eqs.~\eqref{ff2} and \eqref{ff3}, $\Gamma$ is the half-width at half maximum (HWHM) of the cavity, and $2\gamma$ is the spontaneous emission rate of the atom in free space.
\subsection{Single-photon lasers in the dressed-state basis}
Now, let's review the conditions of these lasers using a new basis. 
In the dressed state basis, instead of using the bare states $\ket{0}_\mu$ and $\ket{1}_\mu$, the states $\ket{+}_\mu$ and $\ket{-}_\mu$ can be used, where
\begin{eqnarray*}
	\ket{+}_\mu&=&\cos\alpha\ket{1}_\mu+\sin\alpha\ket{0}_\mu,\\
	\ket{-}_\mu&=&-\sin\alpha\ket{1}_\mu+\cos\alpha\ket{0}_\mu,
	\end{eqnarray*}
    which $\mu$ represents the quantum system coupled to the cavity mode \cite{1992PhRvA.45.2057Z}.
The angle $\alpha$, which lies in the interval $[0,\pi/2]$, is defined as: $\Omega = \Omega^\prime \sin 2\alpha$, and 
$\Delta_1 = \Omega^\prime \cos 2\alpha$.

The effective Rabi frequency,
$\Omega^\prime = (\Omega^2 + \Delta_1^2)^{1/2}$,
is equal to the energy splitting of the dressed states.
In order to obtain the dressed-state Hamiltonian $\mathit{H^\prime}$, we use the unitary transformation $U$:
\begin{equation}\label{key}
\mathit{U}=\prod \exp(i\alpha\sigma_{2\mu})
\end{equation}
In this case, the transformed Hamiltonian is given by
\[
H^\prime = U H U^{-1}.
\]
It can be shown that
\begin{widetext}
    \begin{eqnarray}\label{ff4}
	H^\prime &=& \sum_{\mu}\frac{\Omega^\prime}{2} \sigma_{3\mu}+\Delta_2 a^\dagger a\notag
    +\sum_{\mu} \frac{g_\mu^\ast}{4} a^\dagger[(1+\cos2\alpha)\sigma_\mu-(1-\cos2\alpha)\sigma_\mu^\dagger  +(\sin2\alpha)    \sigma_{3\mu}]\notag\\
	&+&\sum_{\mu}\frac{\mathit{g}_\mu}{4}a[(1+\cos2\alpha)\sigma_\mu^\dagger-(1-\cos2\alpha) \sigma_\mu+(\sin2\alpha) \sigma_{3\mu}].
	\end{eqnarray}
\end{widetext}

    The semiclassical laser equations are obtained in the following form \cite{PhysRevB.1.15, Shakir:84}:
\begin{eqnarray}
\dot{S}&=&-(\gamma_1+i\Omega^\prime)S+i\lambda_1S_3 a\label{eq13}\\
\dot{S}_3&=&-\gamma_2(S_3-\bar{S_3})+2i\gamma_1(a^\ast S-S^\ast a)\label{ll1}\\
\dot{a}&=&-(\Gamma+i\Delta_2)a-i\lambda_1S\label{ll2}
\end{eqnarray}
Here, $S_3$ is the macroscopic population inversion of the dressed states, defined as
\begin{equation*}
S_3=\sum_{\mu}\sigma_{3\mu}.
\end{equation*}
The steady-state value of the dressed-state population inversion,
$\bar{S_3}$, is given by
\begin{equation}\label{gg}
\bar{S_3}=\frac{-2N\cos 2\alpha}{1+\cos^2\alpha}.
\end{equation}
For negative detuning, $\alpha$ is greater than $\pi/4$. As a result, the upper-dressed state has a larger population than the lower-dressed state. Consequently, $\bar{S_3}$ becomes positive.

The resulting equation for the steady-state value of the cavity field amplitude is
\begin{equation}\label{eqq3}
\mathit{a}=\frac{\lambda_1^2 S_3 a}{[\Gamma+i(\Delta_2-\Delta_L)][\gamma_1+i(\Omega^\prime-\Delta_L)]},
\end{equation}
where
\begin{equation}\label{star}
\Delta_L=\frac{\gamma_1 \Delta_2+\Gamma \Omega^\prime}{\gamma_1+\Gamma}.
\end{equation}
Above the threshold, the laser operates in a stable regime. The following condition
    \begin{equation}\label{eqq555}
	1\geq\!-\frac{Ng^2\cos2(\alpha)(1\!+\!\cos2\alpha)^2}{4\gamma\Gamma(2\!+\!\sin^2 2\alpha)(1\!+\!\cos^2 2\alpha)}\!\!\times\!\!  \bigg[1\!+\!\frac{(\Delta_2\!-\!\Omega^\prime)^2}{(\gamma_1\!+\!\Gamma)^2}\bigg]^{-1},
	\end{equation}
    is the condition governing the operation of the laser. The dynamical process presented in this section characterizes the atom–cavity interaction. 

%----------------------------------------------------------------------------------------
%----------------------------------------------------------------------------------------------------
\section{Calculation of the Resonance Fluorescence Spectrum of a Three-Level Quantum Dot in a Cavity under Incoherent Pumping}\label{sec:calculation}

The resonance fluorescence spectrum of the three-level system is significant. We assume that the connection of the excited levels $\ket{1}$ and $\ket{2}$ to the ground level $\ket{0}$ occurs via electric dipole transitions \cite{ficek2005quantum}.  
The cosine of the angle between these two dipole moments controls quantum interference between the two transition paths \cite{ficek2005quantum}.  
We define this parameter $\beta$ as
\begin{equation}
\beta = \frac{\mathbf{d_{10}}\cdot\mathbf{d_{20}}}{|\mathbf{d_{10}}||\mathbf{d_{20}}|},
\end{equation}
where $\mathbf{d_{10}}$ is the dipole moment of the transition between $\ket{1}$ and $\ket{0}$, and $\mathbf{d_{20}}$ is the dipole moment of the transition between $\ket{2}$ and $\ket{0}$. It is obvious that if the electric dipole moments are parallel, quantum interference is maximal, and if they are orthogonal, quantum interference is minimal.
Using the semiclassical Bloch equations, the resonance fluorescence spectrum of the three-level system was studied in Ref. \cite{ficek2005quantum}. The effects of quantum interference can be seen as a triple peak.

We model the three-level system as two excitonic subsystems: the first with excited state $\ket{1}$ and a ground state $\ket{0}$, and the second with excited state $\ket{2}$ and ground state $\ket{0}$.
The evolution of a three-level atom in strong coupling with a single-mode microcavity is described by the Jaynes-Cummings Hamiltonian \cite{1443594, PhysRevB.84.195313}
\begin{equation}\label{shagh}
H = \omega_a a^\dagger a + \sum_{i=1}^{2} \omega_{\sigma_i} \sigma_i^\dagger \sigma_i + \sum_{i=1}^{2} g_i(a^\dagger \sigma_i + a \sigma_i^\dagger),
\end{equation}
 where $a$ is the cavity mode annihilation operator, tuned to energy $\omega_a$, and $\sigma_1$ and $\sigma_2$ are the annihilation operators for the first and second excitons with energies $\omega_{\sigma_1}$ and $\omega_{\sigma_2}$. Also $g_1$ is the coupling constant between cavity mode $\omega_a$ and $\omega_{\sigma_1}$, and $g_2$ is the coupling constant between cavity mode $\omega_a$ and $\omega_{\sigma_2}$.  

The Lindblad equation for this system is \cite{PhysRevB.84.195313}:
\begin{widetext}
\begin{eqnarray}\label{shagh1}
\partial_t \rho &=& i[\rho, H] + \frac{\gamma_a}{2}(2a\rho a^\dagger - a^\dagger a \rho - \rho a^\dagger a) + \sum_{i=1,2} \frac{\gamma_i}{2} (2 \sigma_i \rho \sigma_i^\dagger - \sigma_i^\dagger \sigma_i \rho - \rho \sigma_i^\dagger \sigma_i) \notag \\
&+& \frac{\gamma_{12}}{2}(2 \sigma_1 \rho \sigma_2^\dagger - \sigma_2^\dagger \sigma_1 \rho - \rho \sigma_2^\dagger \sigma_1) + \frac{\gamma_{12}}{2}(2 \sigma_2 \rho \sigma_1^\dagger - \sigma_1^\dagger \sigma_2 \rho - \rho \sigma_1^\dagger \sigma_2) \notag \\
&+& \frac{P_a}{2}(2 a^\dagger \rho a - a a^\dagger \rho - \rho a a^\dagger) + \sum_{i=1,2} \frac{P_i}{2}(2 \sigma_i^\dagger \rho \sigma_i - \sigma_i \sigma_i^\dagger \rho - \rho \sigma_i \sigma_i^\dagger).
\end{eqnarray}
\end{widetext}

Here, $\rho$ is the density matrix of the total quantum dot–cavity system. $\gamma_{12} = \beta \sqrt{\gamma_1 \gamma_2}$, and $P_a$ and $P_i$ are the pumping rates. The pumping originates from the coupling of the whole system to an electron–hole reservoir. Exciton generation in the quantum dot inside the microcavity occurs via annihilation of an electron–hole pair in the external system and creation of a pair in the quantum dot, mediated by phonons \cite{delvalle2009microcavity}. This pumping is represented by $P_i$ in Eq. \eqref{shagh1}. Another type of pumping, represented by $P_a$, arises from nearby spectator quantum dots weakly interacting with the cavity \cite{delvalle2009microcavity, Reitzenstein:06}.

We can obtain the expectation values of operators of the form
${a^\dagger}^m{a}^n{\sigma^\dagger_1}^{\mu_1}\sigma_1^{\nu_1}{\sigma^\dagger_2}^{\mu_2}\sigma_2^{\nu_2}$.
The quantity of interest is the light spectrum, which is defined in \cite{PhysRevB.79.235326} as
\begin{widetext}
\begin{equation}\label{ley1}
S_c(\omega)=\frac{1}{\pi \int_{0}^{\infty}\langle a^\dagger(t) c(t)\rangle dt}\,\mathrm{Re}\int_{0}^{\infty}\int_{0}^{\infty}\langle a^\dagger(t)c(t+t^\prime)\rangle e^{i\omega t^\prime}d t^\prime dt,\quad  c=a, \sigma_1, \sigma_2.
\end{equation}
\end{widetext}
The normalization factor is defined such that $$\int S_a(\omega)d\omega=1.$$

Let's calculate $\langle c^{\dagger}(t)c(t+t^\prime)\rangle$ from the differential equation of motion
\begin{equation}\label{shagh3}
 \frac{\partial\mathbf{u}_c(t,t+t^\prime)}{\partial t^\prime}=\mathbf{T}\mathbf{u}_c(t,t+t^\prime).
\end{equation}  
For $c=a$, we have
\begin{eqnarray}\label{khar}
\mathbf{u}_a(t,t+t^\prime)=\begin{pmatrix}
\langle a^\dagger(t)a(t+t^\prime)\rangle\\
\langle a^\dagger(t)\sigma_1(t+t^\prime)\rangle\\
\langle a^\dagger(t)\sigma_2(t+t^\prime)\rangle\\
\vdots
\end{pmatrix},
\end{eqnarray}
with the initial value of this vector $\mathbf{u}_a(t,t)$. There are more elements included in the vector in Eq. (\ref{khar}). Some of these elements were included in the vector calculated in Appendix \ref{sec:AppendixA}.
The first term we calculate is $\langle a^\dagger(t)\frac{\partial a(t+t^\prime)}{\partial t^\prime}\rangle$, noting that $\partial_t^\prime a^\dagger(t)=0$.
We can show that
\begin{widetext}
\begin{eqnarray}\label{armin1}
\partial_t^\prime\langle a^\dagger(t)a(t+t^\prime)\rangle 
=  (-i\omega_a - \frac{\Gamma_a}{2}) \langle a^\dagger(t)a(t+t^\prime)\rangle 
 - i g_1 \langle a^\dagger(t)\sigma_1(t+t^\prime)\rangle - i g_2 \langle a^\dagger(t)\sigma_2(t+t^\prime)\rangle,\notag\\
\end{eqnarray}
\end{widetext}
where $\Gamma_a = \gamma_a - P_a$ and $\Gamma_{\sigma_i} = \gamma_i + P_i$.

Next, we calculate $\partial_t^\prime \langle a^\dagger(t)\sigma_1(t+t^\prime)\rangle$, 
\begin{widetext}
\begin{eqnarray}\label{shagh6}
\partial_t^\prime \langle a^\dagger(t)\sigma_1(t+t^\prime)\rangle 
&=&  - i g_1 \langle a^\dagger(t)a(t+t^\prime)\rangle 
+ (-i \omega_{\sigma_1} - \frac{\Gamma_{\sigma_1}}{2}) \langle a^\dagger(t)\sigma_1(t+t^\prime)\rangle\notag\\
&-& \frac{\gamma_{12}}{2} \langle a^\dagger(t)\sigma_2(t+t^\prime)\rangle 
 + 2 i g_1 \langle a^\dagger(t) (a \sigma_1^\dagger \sigma_1)|_{(t+t^\prime)} \rangle
+ \gamma_{12} \langle a^\dagger(t) (\sigma_1^\dagger \sigma_1 \sigma_2)|_{(t+t^\prime)} \rangle.
\end{eqnarray}
\end{widetext}
which can be used to calculate $S_{\sigma_1}(\omega)$, and similarly for $S_{\sigma_2}(\omega)$, it is necessary to find $\partial_t^\prime \langle a^\dagger(t)\sigma_2(t+t^\prime)\rangle$
\begin{widetext}
\begin{eqnarray}\label{shagh7}
\partial_t^\prime \langle a^\dagger(t)\sigma_2(t+t^\prime)\rangle
&=& - i g_2 \langle a^\dagger(t)a(t+t^\prime)\rangle 
+ (-i\omega_{\sigma_2} - \frac{\Gamma_{\sigma_2}}{2}) \langle a^\dagger(t)\sigma_2(t+t^\prime)\rangle \notag\\
& -& \frac{\gamma_{12}}{2} \langle a^\dagger(t)\sigma_1(t+t^\prime)\rangle
+ 2 i g_2 \langle a^\dagger(t) (a \sigma_2^\dagger \sigma_2)|_{(t+t^\prime)} \rangle 
 + \gamma_{12} \langle a^\dagger(t) (\sigma_1 \sigma_2^\dagger \sigma_2)|_{(t+t^\prime)} \rangle.
\end{eqnarray}
\end{widetext}
Consequently, the matrix $\mathbf{T}$ is
\begin{eqnarray}\label{cal1}
\mathbf{T}=\begin{pmatrix}
-i\omega_a-\frac{\Gamma_a}{2} & -i g_1 & -i g_2\\
-i g_1 & -i\omega_{\sigma_1}-\frac{\Gamma_{\sigma_1}}{2} & -\frac{\gamma_{12}}{2}\\
-i g_2 & -\frac{\gamma_{12}}{2} & -i\omega_{\sigma_2}-\frac{\Gamma_{\sigma_2}}{2}
\end{pmatrix}.
\end{eqnarray}

To obtain the elements of $\mathbf{u}$ in Eq.~\eqref{shagh3}, we use
\begin{equation}\label{goo1}
\mathbf{u}_a(t,t+t^\prime) = \mathbf{V} e^{-\mathbf{D}t^\prime} \mathbf{V}^{-1} \mathbf{u}_a(t,t),
\end{equation}
where $\mathbf{V}$ is the normalized eigenvector matrix and $\mathbf{D}$ is the eigenvalue matrix, such that $\mathbf{T} = \mathbf{V} \left(-\mathbf{D}\right) \mathbf{V}^{-1}$ \cite{PhysRevB.79.235326}.
Let's define these matrices as
\begin{eqnarray}
\mathbf{V} &=& \begin{pmatrix} a_1 & a_2 & a_3\\ a_4 & a_5 & a_6\\ a_7 & a_8 & a_9 \end{pmatrix}, \quad
\mathbf{V}^{-1} = \begin{pmatrix} b_1 & b_2 & b_3\\ b_4 & b_5 & b_6\\ b_7 & b_8 & b_9 \end{pmatrix}, \\
e^{-\mathbf{D}t^\prime} &=& \begin{pmatrix} e^{-\lambda_1 t^\prime} & 0 & 0\\ 0 & e^{-\lambda_2 t^\prime} & 0\\ 0 & 0 & e^{-\lambda_3 t^\prime} \end{pmatrix}.
\end{eqnarray}
Using the method described in Ref.~\cite{PhysRevB.79.235325}, we take time derivatives of the matrix elements and then find their steady-state values.
The steady-state values are shown in Appendix \ref{sec:AppendixA}.
Next, we calculate $\mathbf{u}_a(t,t)$. 
Let's define
\begin{eqnarray*}
\begin{pmatrix}
\langle a^\dagger(t) a(t)\rangle\\
\langle a^\dagger(t) \sigma_1(t)\rangle\\
\langle a^\dagger(t) \sigma_2(t)\rangle
\end{pmatrix}=
\begin{pmatrix} n_a(t)\\ n_{a\sigma_1}(t)\\ n_{a\sigma_2}(t)\end{pmatrix}
\end{eqnarray*}
for simplicity.
Then Eq.~\eqref{goo1} becomes
\begin{widetext}
\begin{eqnarray*}
\begin{pmatrix}
\langle a^\dagger(t) a(t+t^\prime) \rangle\\
\langle a^\dagger(t) \sigma_1(t+t^\prime)\rangle\\
\langle a^\dagger(t) \sigma_2(t+t^\prime)\rangle
\end{pmatrix}
\!=\!
\begin{pmatrix}
e^{-\lambda_1 t^\prime} a_1 A_1(t) + e^{-\lambda_2 t^\prime} a_2 A_2(t) + e^{-\lambda_3 t^\prime} a_3 A_3(t) \\
e^{-\lambda_1 t^\prime} a_4 A_1(t) + e^{-\lambda_2 t^\prime} a_5 A_2(t) + e^{-\lambda_3 t^\prime} a_6 A_3(t) \\
e^{-\lambda_1 t^\prime} a_7 A_1(t) + e^{-\lambda_2 t^\prime} a_8 A_2(t) + e^{-\lambda_3 t^\prime} a_9 A_3(t)
\end{pmatrix},
\end{eqnarray*}
\end{widetext}
where
\begin{eqnarray}\label{hhh1}
A_1(t) &= b_1 n_a(t) + b_2 n_{a\sigma_1}(t) + b_3 n_{a\sigma_2}(t),\notag\\
A_2(t) &= b_4 n_a(t) + b_5 n_{a\sigma_1}(t) + b_6 n_{a\sigma_2}(t),\notag\\
A_3(t) &= b_7 n_a(t) + b_8 n_{a\sigma_1}(t) + b_9 n_{a\sigma_2}(t).\notag\\
\end{eqnarray}

The radiation spectrum is then obtained using Eq.~\eqref{ley1} \cite{PhysRevB.79.235325}.
The steady-state spectrum is
\begin{eqnarray*}
S_a^{SS}(\omega) = \frac{1}{\pi} \frac{1}{n_a^{SS}} \lim_{t\to\infty} \mathrm{Re} \int_0^\infty \langle a^\dagger(t) a(t+t^\prime) \rangle e^{i\omega t^\prime} dt^\prime,
\end{eqnarray*}
where $n_a^{SS}$ is the steady-state population $\langle a^\dagger(t) a(t) \rangle$. 
The steady-state values $A_1^{SS}, A_2^{SS}, A_3^{SS}$ in Eq.~\eqref{hhh1} are
\begin{eqnarray}
A_1^{SS} &= b_1 n_a^{SS} + b_2 n_{a\sigma_1}^{SS} + b_3 n_{a\sigma_2}^{SS},\notag\\
A_2^{SS} &= b_4 n_a^{SS} + b_5 n_{a\sigma_1}^{SS} + b_6 n_{a\sigma_2}^{SS},\notag\\
A_3^{SS} &= b_7 n_a^{SS} + b_8 n_{a\sigma_1}^{SS} + b_9 n_{a\sigma_2}^{SS}.\notag\\
\end{eqnarray}

Note that 
$$\int_0^\infty e^{-\lambda_i t^\prime} e^{i\omega t^\prime} dt^\prime = \frac{1}{\lambda_i - i \omega} = \frac{1}{\mathrm{Re}(\lambda_i) - i[\omega - \mathrm{Im}(\lambda_i)]},$$ and 
$$\mathrm{Re} \int_0^\infty e^{-\lambda_i t^\prime} e^{i\omega t^\prime} dt^\prime = \frac{\mathrm{Re}(\lambda_i)}{[\omega - \mathrm{Im}(\lambda_i)]^2 + [\mathrm{Re}(\lambda)]^2}.$$
Here, $\lambda_i$s are the eigenvalues of the matrix $\mathbf{T}$ in Eq. \ref{cal1}.
Now, with some calculations, we can get
\begin{widetext}
\begin{eqnarray}\label{eq:s_a}
S_a^{SS}(\omega) &=& \frac{1}{\pi} \frac{1}{n_a^{SS}} \bigg[
\frac{a_1 \mathrm{Re}(\lambda_1) A_1^{SS}}{[\omega - \mathrm{Im}(\lambda_1)]^2 + [\mathrm{Re}(\lambda_1)]^2} 
+
\frac{a_2 \mathrm{Re}(\lambda_2) A_2^{SS}}{[\omega - \mathrm{Im}(\lambda_2)]^2 + [\mathrm{Re}(\lambda_2)]^2} 
 + \frac{a_3 \mathrm{Re}(\lambda_3) A_3^{SS}}{[\omega - \mathrm{Im}(\lambda_3)]^2 + [\mathrm{Re}(\lambda_3)]^2}
\bigg].
\end{eqnarray}
\end{widetext}

Similarly, for $S_{\sigma_1}^{SS}(\omega)$ and $S_{\sigma_2}^{SS}(\omega)$ 
\begin{widetext}
\begin{eqnarray}
  			S_{\sigma_1}^{SS}(\omega)&=&\frac{1}{\pi}\frac{1}{n_{\sigma_1}^{SS}}\bigg[
  			\frac{a_4\mathrm{Re}(\lambda_1)A_1^{SS}}{{[\omega-\Im(\lambda_1)]^2+[\mathrm{Re}(\lambda_1)]^2}}
            +\frac{a_5\mathrm{Re}(\lambda_2)A_2^{SS}}{{[\omega-\Im(\lambda_2)]^2+[\mathrm{Re}(\lambda_2)]^2}}
  			+\frac{a_6\mathrm{Re}(\lambda_3)A_3^{SS}}{{[\omega-\Im(\lambda_3)]^2+[\mathrm{Re}(\lambda_3)]^2}}
  			\bigg]\label{eq:ss1}\\
  			S_{\sigma_2}^{SS}(\omega)&=&\frac{1}{\pi}\frac{1}{n_{\sigma_2}^{SS}}\bigg[
  			\frac{a_7\mathrm{Re}(\lambda_1)A_1^{SS}}{{[\omega-\Im(\lambda_1)]^2+[\mathrm{Re}(\lambda_1)]^2}}
            +\frac{a_8\mathrm{Re}(\lambda_2)A_2^{SS}}{{[\omega-\Im(\lambda_2)]^2+[\mathrm{Re}(\lambda_2)]^2}}
  			+\frac{a_9\mathrm{Re}(\lambda_3)A_3^{SS}}{{[\omega-\Im(\lambda_3)]^2+[\mathrm{Re}(\lambda_3)]^2}}
  			\bigg]\label{eq:ss2}.
  			\end{eqnarray}
\end{widetext}

\begin{figure*}[ht]
    \centering
    \begin{subfigure}[t]{0.49\textwidth}
\includegraphics[width=0.6\textwidth]{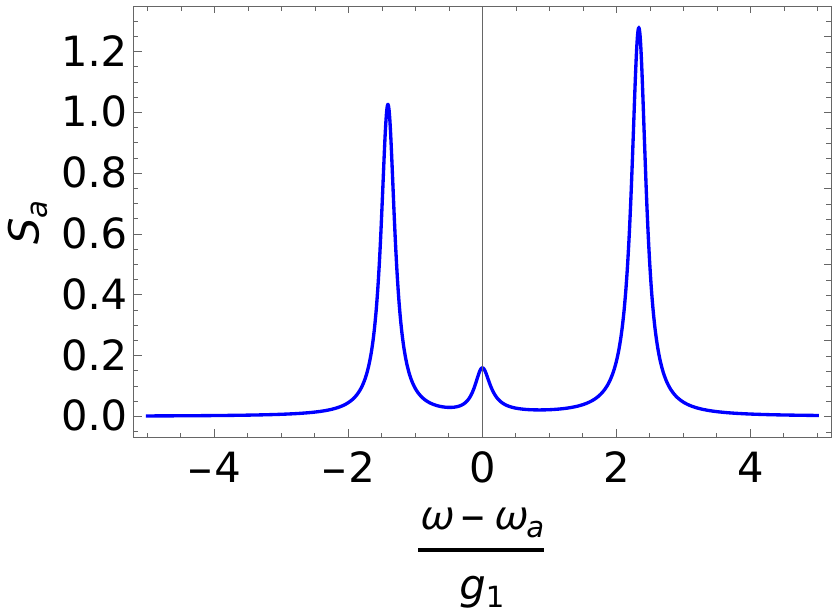}
         \caption{} 
    \end{subfigure}
    \begin{subfigure}[t]{0.49\textwidth}
    \includegraphics[width=0.6\textwidth]{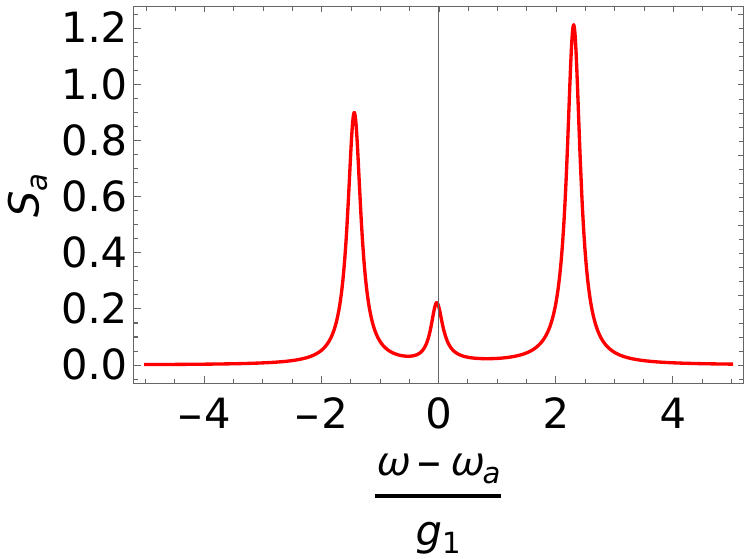}
         \caption{}  
    \end{subfigure}
    \begin{subfigure}[t]{0.49\textwidth}
    \includegraphics[width=0.6\textwidth]{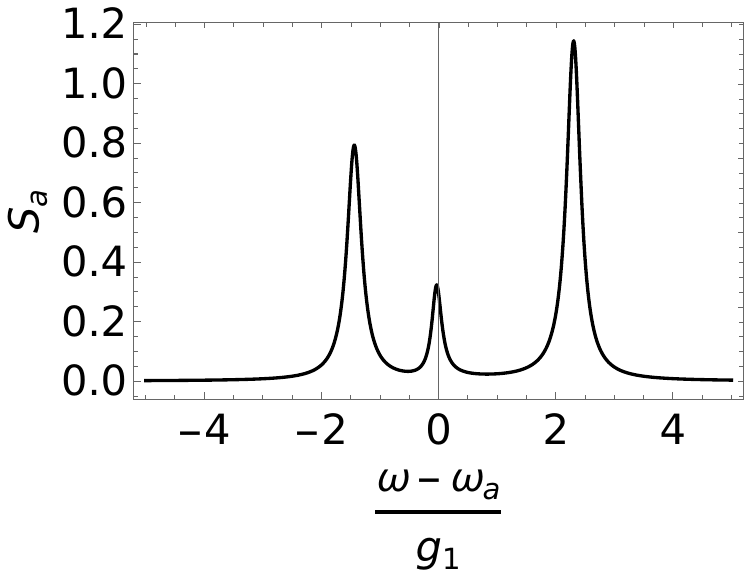}
    \caption{} 
    \end{subfigure}
    \begin{subfigure}[t]{0.49\textwidth}
\includegraphics[width=0.6\textwidth]{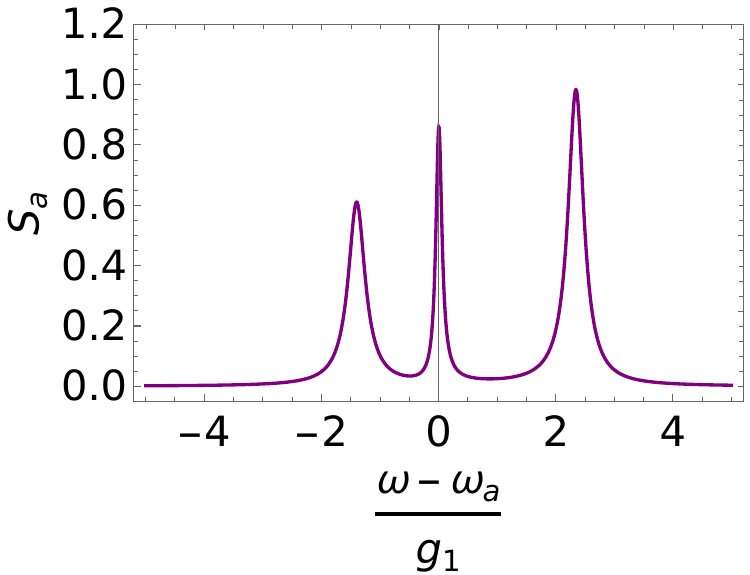}
\caption{} 
    \end{subfigure}
    \caption{
        $S_a$ 
        as a function of 
        $\frac{\omega-\omega_a}{g_1}$. 
        (a) $\beta=0$, 
        (b) $\beta=\frac14$, 
        (c) $\beta=\frac12$, 
        (d) $\beta=1$. 
        The parameter values are:
        $\gamma_a=0.3g_1$, 
        $\gamma_1=0.15g_1$, 
        $\gamma_2=0.2g_1$, 
        $g_2=1.5g_1$, 
        $\Delta_1=g_1$, 
        and 
        $\Delta_2=0$. 
        Additionally, 
        $P_a=0.1g_1$, 
        $P_1=0.1g_1$, 
        and 
        $P_2=0.1g_2$.
    }\label{fig:Sa1}
\end{figure*}

Now let's proceed to the numerical analysis of the spectra. In Figure~\ref{fig:Sa1}, 
$S_a(\omega)$ 
is plotted for different values of 
$\beta$ 
as a function of 
$\frac{\omega - \omega_a}{g_1}$. 
$S_a$ represents the cavity emission~\cite{PhysRevB.79.235326}. 
In all plots, three peaks can be observed. The central peak, located at the origin, occurs at the frequency 
$\omega_a$ 
(since the plots are drawn in terms of 
$\omega - \omega_a$, this peak is shifted and centered on the horizontal axis). 
It can be seen that increasing the quantum interference, i.e., 
$\beta$, 
leads to an increase in the height of the central peak, which corresponds to an increase in the emission intensity at the frequency 
$\omega_a$. This increase reaches its maximum at 
$\beta = 1$, 
which is clearly visible.
The change in height of the side peaks, when varying 
$\beta$ from $0$ to $\frac{1}{4}$ and $\frac{1}{2}$, 
is relatively small (especially for the peak located on the positive side of the axis). 
However, when 
$\beta$ 
is increased from $0$ to $1$, the decrease in height of both side peaks becomes clearly noticeable. 
Another point is that with increasing 
$\beta$, 
the central peak becomes narrower, giving a more defined peak at the desired frequency.

\begin{figure*}[ht]
    \centering
    \begin{subfigure}[t]{0.49\textwidth}
\includegraphics[width=0.6\textwidth]{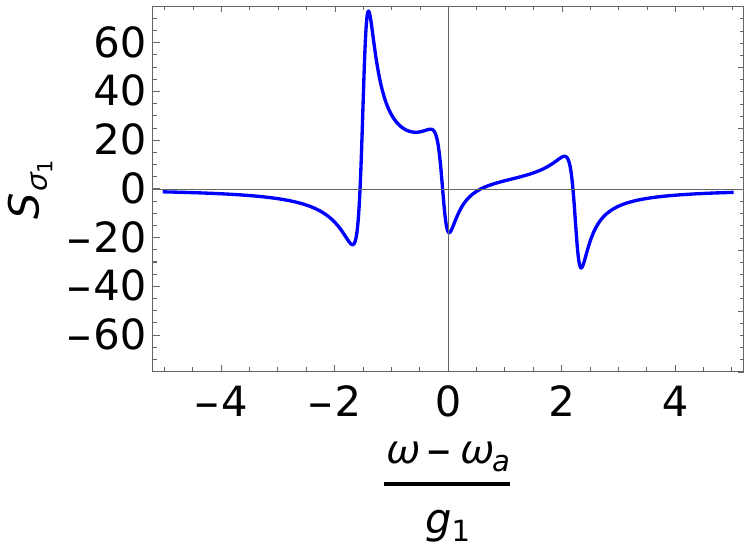}
         \caption{} 
    \end{subfigure}
    \begin{subfigure}[t]{0.49\textwidth}
    \includegraphics[width=0.6\textwidth]{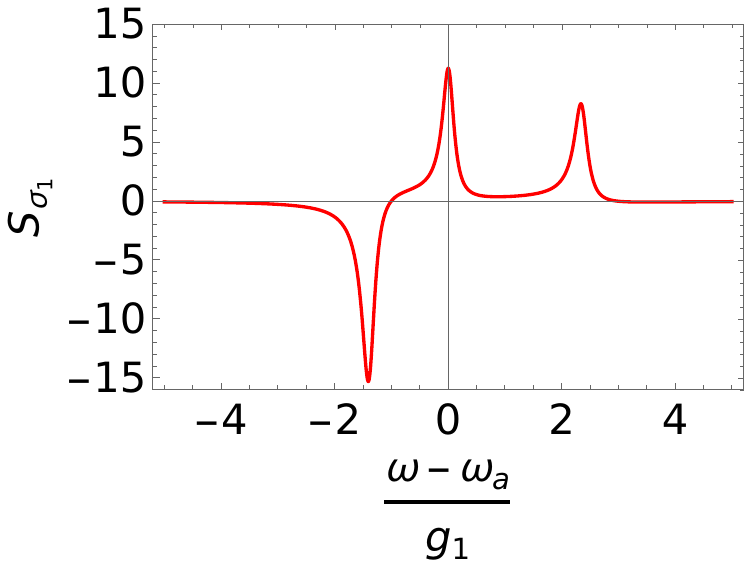}
         \caption{}  
    \end{subfigure}
    \begin{subfigure}[t]{0.49\textwidth}
    \includegraphics[width=0.6\textwidth]{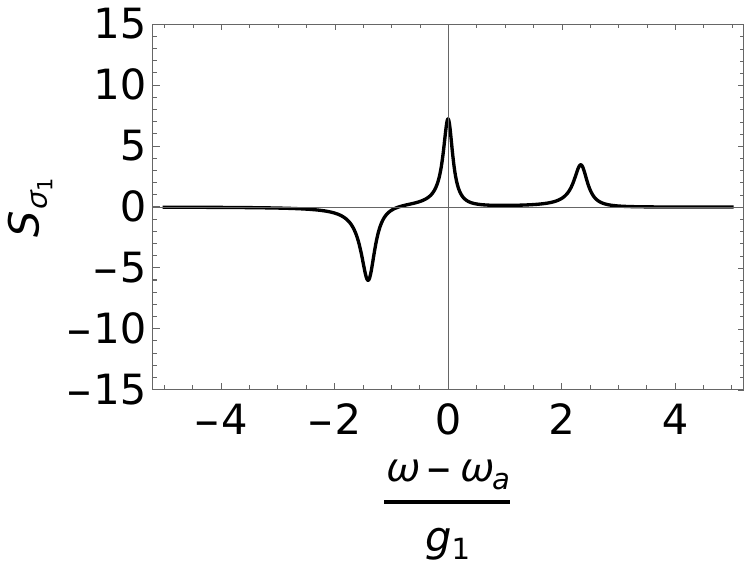}
    \caption{} 
    \end{subfigure}
    \begin{subfigure}[t]{0.49\textwidth}
\includegraphics[width=0.6\textwidth]{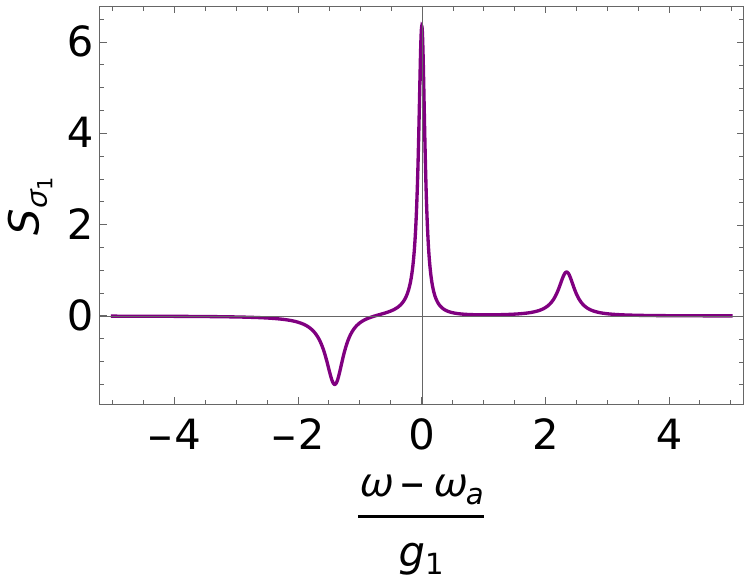}
\caption{} 
    \end{subfigure}
    \caption{
        $S_{\sigma_1}$ as a function of $\frac{\omega-\omega_a}{g_1}$.
(a) $\beta = 0$
(b) $\beta = \frac{1}{4}$
(c) $\beta = \frac{1}{2}$
(d) $\beta = 1$.
The parameter values are as follows:
$\gamma_a = 0.3g_1$,
$\gamma_1 = 0.15g_1$,
$\gamma_2 = 0.2g_1$,
$g_2 = 1.5g_1$,
$\Delta_1 = g_1$,
and $\Delta_2 = 0$.
Also,
$P_a = 0.1g_1$,
$P_1 = 0.1g_1$,
and
$P_2 = 0.1g_2$.
    }\label{fig:s1}
\end{figure*}

\begin{figure*}[ht]
     \centering
    \begin{subfigure}[t]{0.49\textwidth}
\includegraphics[width=0.6\textwidth]{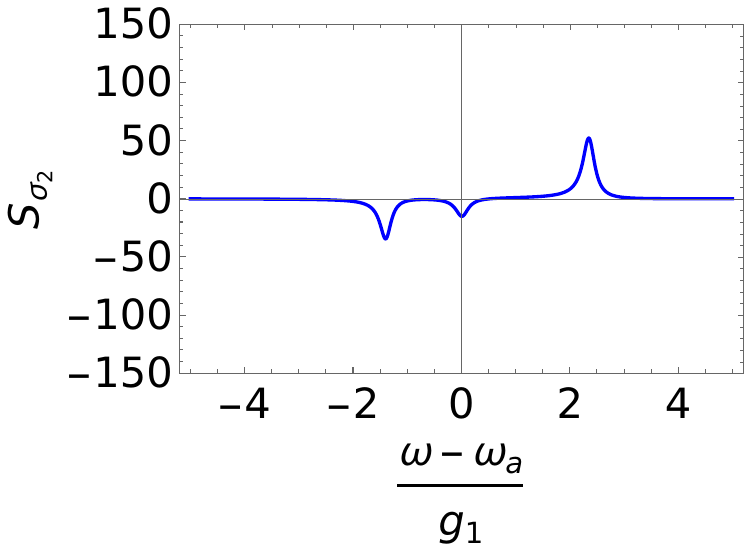}
         \caption{} 
    \end{subfigure}
    \begin{subfigure}[t]{0.49\textwidth}
    \includegraphics[width=0.6\textwidth]{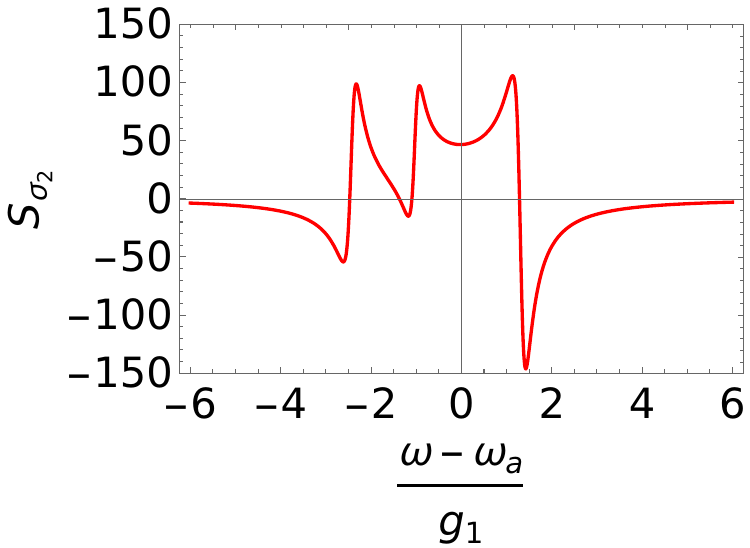}
         \caption{}  
    \end{subfigure}
    \begin{subfigure}[t]{0.49\textwidth}
    \includegraphics[width=0.6\textwidth]{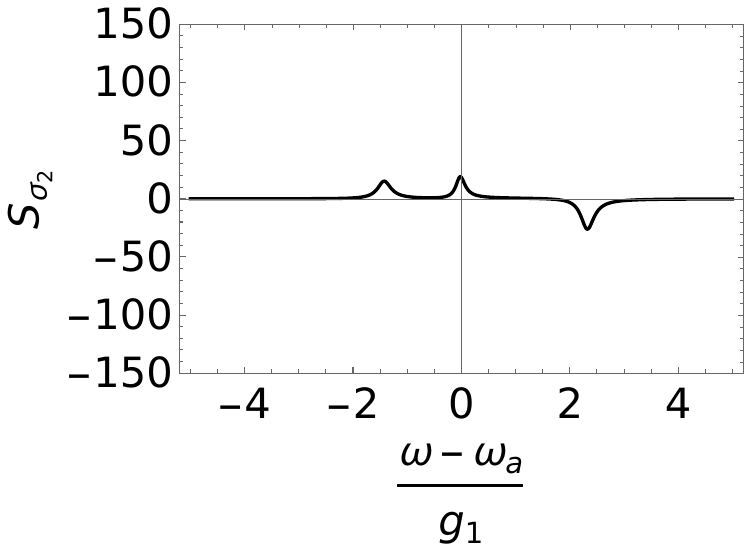}
    \caption{} 
    \end{subfigure}
    \begin{subfigure}[t]{0.49\textwidth}
\includegraphics[width=0.6\textwidth]{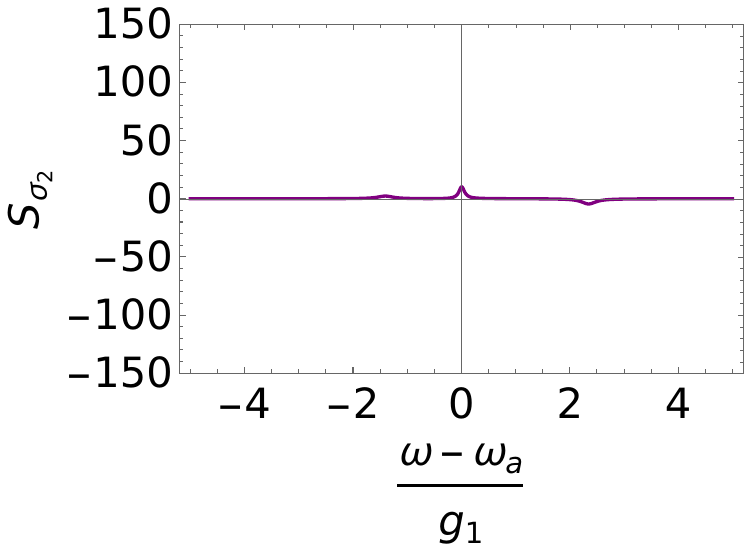}
\caption{} 
    \end{subfigure}
    \caption{
        $S_{\sigma_2}$ as a function of $\frac{\omega-\omega_a}{g_1}$.
(a) $P_a = 0.01g_1$,
(b) $P_a = 0.1g_1$,
(c) $P_a = 1g_1$,
(d) $P_a = 10g_1$.
The parameter values are:
$\gamma_a = 0.3g_1$,
$\gamma_1 = 0.15g_1$,
$\gamma_2 = 0.2g_1$,
$g_2 = 1.5g_1$,
$\Delta_1 = g_1$,
and $\Delta_2 = 0$.
The blue curve corresponds to $\beta = 0$, while the red curve corresponds to $\beta = 1$.}\label{fig:s2}
\end{figure*}

\begin{figure*}[ht]
     \centering
    \begin{subfigure}[t]{0.49\textwidth}
\includegraphics[width=0.6\textwidth]{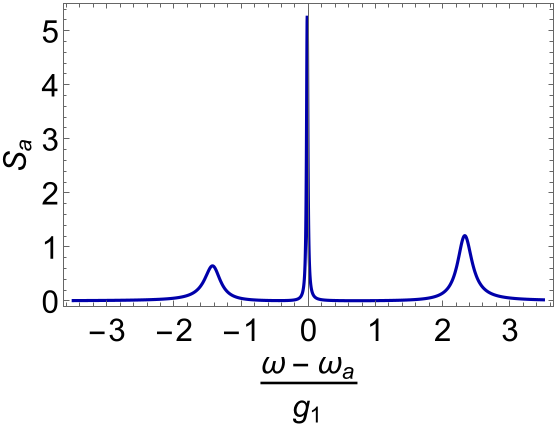}
         \caption{} 
    \end{subfigure}
    \begin{subfigure}[t]{0.49\textwidth}
    \includegraphics[width=0.6\textwidth]{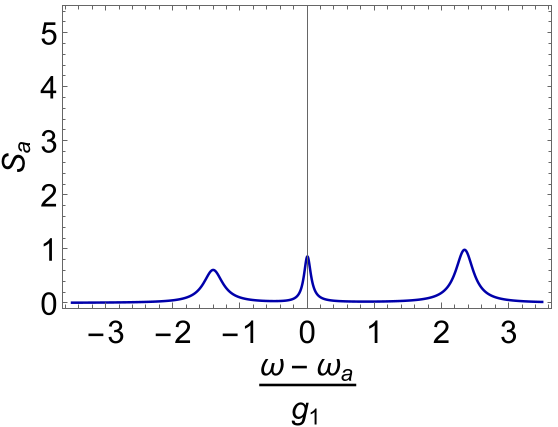}
         \caption{}  
    \end{subfigure}
    \begin{subfigure}[t]{0.49\textwidth}
    \includegraphics[width=0.6\textwidth]{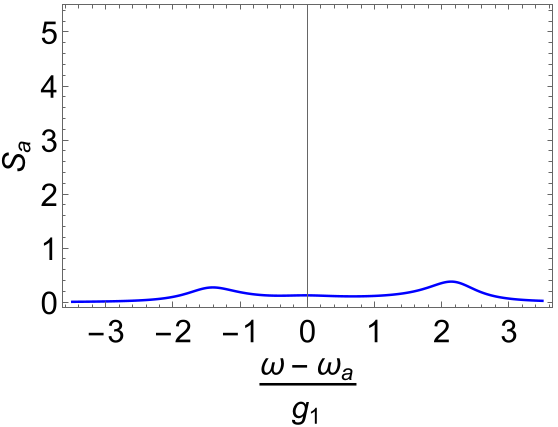}
    \caption{} 
    \end{subfigure}
    \begin{subfigure}[t]{0.49\textwidth}
\includegraphics[width=0.6\textwidth]{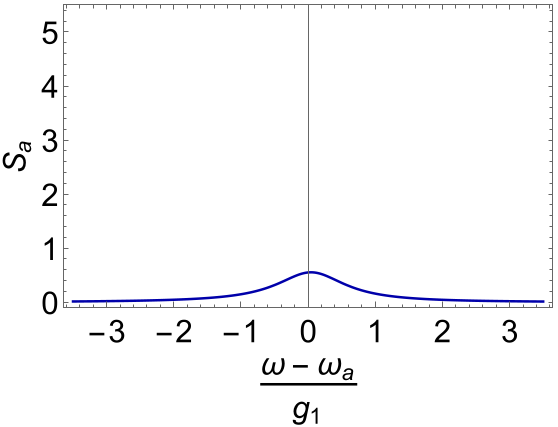}
\caption{} 
    \end{subfigure}
    \caption{$S_a$ as a function of $\frac{\omega-\omega_a}{g_1}$. 
(a) $P_1=0.01g_1$, $P_2=0.01g_2$;
(b) $P_1=0.1g_1$, $P_2=0.1g_2$;
(c) $P_1=1g_1$, $P_2=1g_2$;
(d) $P_1=10g_1$, $P_2=10g_2$.
The parameter values are:
$\gamma_a=0.3g_1$,
$\gamma_1=0.15g_1$,
$\gamma_2=0.2g_1$,
$g_2=1.5g_1$,
$\Delta_1=g_1$,
and $\Delta_2=0$.
The blue curve corresponds to $\beta=0$, and the red curve corresponds to $\beta=1$.
    }\label{fig:sa2}
\end{figure*}

\begin{figure*}[ht]
     \centering
    \begin{subfigure}[t]{0.49\textwidth}
\includegraphics[width=0.6\textwidth]{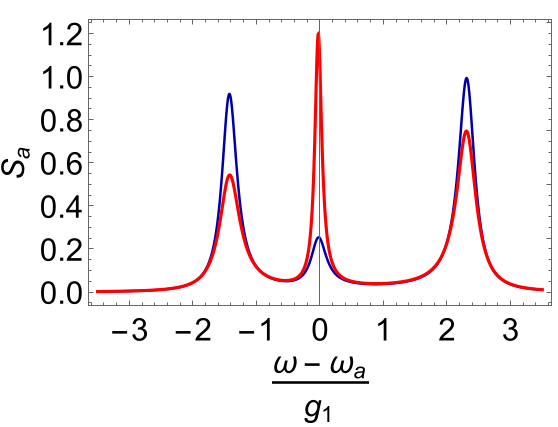}
         \caption{} 
    \end{subfigure}
    \begin{subfigure}[t]{0.49\textwidth}
    \includegraphics[width=0.6\textwidth]{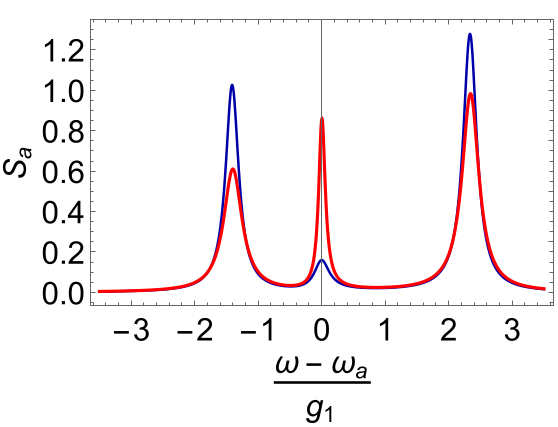}
         \caption{}  
    \end{subfigure}
    \begin{subfigure}[t]{0.49\textwidth}
    \includegraphics[width=0.6\textwidth]{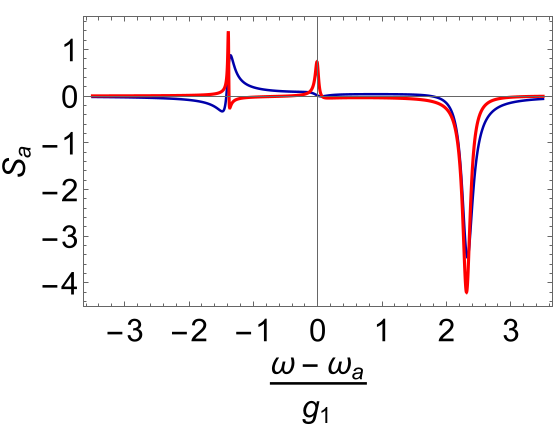}
    \caption{} 
    \end{subfigure}
    \begin{subfigure}[t]{0.49\textwidth}
\includegraphics[width=0.6\textwidth]{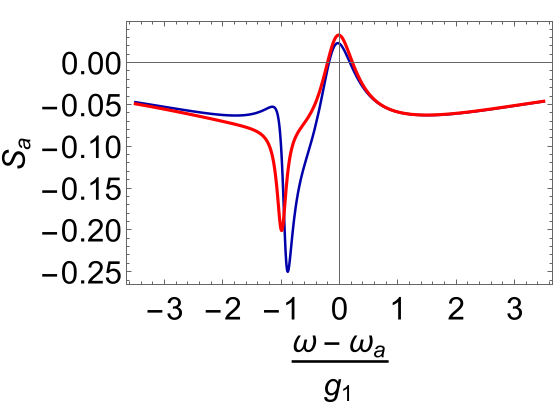}
\caption{} 
    \end{subfigure}
    \caption{$S_a$ as a function of $\frac{\omega-\omega_a}{g_1}$.
(a) $P_a = 0.01g_1$
(b) $P_a = 0.1g_1$
(c) $P_a = 1g_1$
(d) $P_a = 10g_1$.
The parameter values are as follows:
$\gamma_a = 0.3g_1$,
$\gamma_1 = 0.15g_1$,
$\gamma_2 = 0.2g_1$,
$g_2 = 1.5g_1$,
$\Delta_1 = g_1$,
and $\Delta_2 = 0$.
The blue curve corresponds to $\beta = 0$, and the red curve corresponds to $\beta = 1$.
    }\label{fig:sa3}
\end{figure*}
Figure~\ref{fig:s1} shows the behavior of 
$S_{\sigma_1}$ 
as a function of 
$\frac{\omega - \omega_a}{g_1}$. 
Plot (a), drawn for 
$\beta = 0$, 
differs more significantly from the other plots. In the other plots, two peaks and one valley are observed, whereas plot (a) exhibits three valleys and two peaks. 
The width of the central peak decreases with increasing 
$\beta$, 
while the change in the width of the side peak and valley is more gradual. 
Another point is that the peak of 
$S_{\sigma_1}$ 
is located on the positive side of the axis, and its valley is on the negative side. 
In contrast, in Figure~\ref{fig:s2}, we see that the peak of 
$S_{\sigma_2}$ 
is on the negative side and its valley is on the positive side.

In Figure~\ref{fig:s2}, the behavior of 
$S_{\sigma_2}$ 
as a function of 
$\frac{\omega - \omega_a}{g_1}$ 
is plotted. Here again, the plot corresponding to 
$\beta = 0$ 
differs noticeably from the others. When 
$\beta = 0$, 
$S_{\sigma_2}$ 
has two valleys and one peak, but when 
$\beta \neq 0$, 
$S_{\sigma_2}$ 
has two peaks and one valley. 
It is evident that increasing 
$\beta$ 
strongly reduces the size of the peaks and valleys and also causes frequency shifts.

Figure~\ref{fig:sa2} shows the behavior of 
$S_a$ 
as a function of 
$\frac{\omega - \omega_a}{g_1}$ 
for different values of 
$P_1$ 
and 
$P_2$. 
The blue curve corresponds to 
$\beta = 0$ 
and the red curve corresponds to 
$\beta = 1$. 
Examining plot (a), three peaks at different frequencies can be observed, with the central peak higher and narrower than the side peaks. 
Increasing the pump to 
$P_1 = 0.1 g_1$ 
and 
$P_2 = 0.1 g_2$ 
in (b) reduces the central peak height and broadens it, while the side peaks are less affected and only slightly reduced in height. Their width increases slightly as well. 
With a further pump increase in (c), the central peak disappears, and the side peaks decrease more, resulting in a transition from three peaks at different frequencies to two peaks. 
Finally, pump values 
$P_1 = 10 g_1$ 
and 
$P_2 = 10 g_2$ 
create a single central peak in plot (d). 
These changes are similar for both 
$\beta = 0$ 
and 
$\beta = 1$. 
We note that in two-level systems, increasing the exciton pump 
$P_\sigma$ 
produces similar effects~\cite{PhysRevB.79.235326}.

Finally, the effect of changing
$P_a$ 
is analyzed in Figure~\ref{fig:sa3}. 
In plot (a), three peaks at different frequencies are observed, with the central peak higher and narrower than the side peaks. By increasing 
$P_a$ 
in (b), the central peak height decreases and increases the height of the side peaks. No significant change is observed in the peak widths. 
So far, the changes have affected only the height and width of the peaks, and these effects are similar for both values of 
$\beta$. 
In ~\ref{fig:sa3}(c), with 
$P_a = g_1$, 
for 
$\beta = 0$, 
there is one valley and one peak on the negative side of the axis, and one valley on the positive side. 
For 
$\beta = 1$, 
there is one peak and one valley on the negative side, one peak at the origin, and one valley on the positive side. 
A notable feature is the appearance of a valley that was not present in the $S_a$ spectrum previously. 
In  ~\ref{fig:sa3}(d), for both values of 
$\beta$, 
there is one valley on the negative side, one peak at the origin, and one valley on the positive side. 
The only difference between 
$\beta = 0$ 
and 
$\beta = 1$ 
is that 
$\beta = 0$ 
also shows a peak on the negative side. 
The difference in ~\ref{fig:sa3}(d) from the other plots is that, unlike the others, which rapidly approach zero for 
$\abs{\frac{\omega - \omega_a}{g_1}} > 3$, 
here, this occurs at much larger values of 
$\abs{\frac{\omega - \omega_a}{g_1}}$.

\section{Conclusion}

  Aspects of quantum optics related to resonance fluorescence and single-photon sources have been reviewed. We looked into the fundamental properties of the resonance fluorescence spectrum of a two–level atom driven by an external field. This generalizes to any two-level quantum system. In particular, the Mollow triplet structure consists of three spectral peaks located at the atomic transition frequency and at frequencies shifted by the Rabi frequency. It has been shown in several studies that the spectral characteristics of resonance fluorescence can change significantly under incoherent pumping conditions, providing additional possibilities for controlling the emitted radiation.

Furthermore, the paper reviewed the concept of dressed-state quantum systems and their role in the formation of dressed–state lasers. The luminescence spectrum of a quantum dot embedded in a microcavity was also considered, illustrating how cavity quantum electrodynamics can modify light–matter interaction and lead to controllable emission properties. Such systems are of particular interest for the realization of efficient and deterministic single–photon sources.

 Numerical investigations of the emission spectra demonstrated how quantum interference and system parameters influence the spectral structure and intensity distribution of the emitted light, showing that the results are in agreement with the theory.

Overall, the results presented in this paper contribute to a better understanding of resonance fluorescence, photon correlations, and engineered emission in quantum optical systems. These studies are relevant for the development of quantum technologies, including quantum communication, quantum information processing, and solid–state single–photon sources.

\clearpage

\appendix
\section{Steady-state values of $\mathbf{u}_a(t, t)$}\label{sec:AppendixA}

To find the initial values of Eq. (\ref{khar}),  $\mathbf{u}_a(t,t)$ using Eq. (\ref{shagh}), and Eq. (\ref{shagh1}), we need to find the steady-state values of
\begin{widetext}
\begin{align}\label{alaki}
  			&\partial_t \begin{pmatrix}
  				\langle a^\dagger(t) a(t) \rangle\\
  				\langle a^\dagger(t) \sigma_1(t) \rangle\\
  				\langle a(t) \sigma_1^\dagger(t) \rangle\\
  				\langle \sigma_1^\dagger(t) \sigma_1(t) \rangle\\
  				\langle \sigma_1^\dagger(t) \sigma_2(t) \rangle\\
  				\langle a^\dagger(t) \sigma_2(t)\rangle\\
  				\langle a(t) \sigma_2^\dagger(t) \rangle\\
  				\langle \sigma_2^\dagger(t) \sigma_2(t) \rangle\\
  				\langle \sigma_1(t) \sigma_2^\dagger(t) \rangle\\
  				\end{pmatrix}=\begin{pmatrix}
  				P_a\\
  				0\\
  				0\\
  				P_1\\
  				0\\
  			0\\
  				0\\
  				P_2\\
  				0\\
  				\end{pmatrix}\notag\\
  				&+\setlength{\arraycolsep}{2pt}
  				\renewcommand{\arraystretch}{0.8}\tiny\begin{pmatrix}
  				-\Gamma_a & -ig_1 & ig_1 & 0 & 0 & -ig_2 & ig_2 & 0 & 0\\
  				-ig_1 & i(\omega_a\!\!-\!\!\omega_{\sigma_1}\!)\!\!-\!\!\frac{\Gamma_a\!+\!\Gamma_{\sigma_1}}{2} & 0 & ig_1 & 0 & -\frac{\gamma_{12}}{2} & 0 & 0 & ig_2\\
  				ig_1 & 0 & i(\omega_{\sigma_1}\!\!\!-\!\!\omega_a\!)\!\!-\!\!\frac{\Gamma_a\!+\!\Gamma_{\sigma_1}}{2} & -ig_1 & -ig_2 & 0 & -\frac{\gamma_{12}}{2} & 0 & 0\\
  				0 & ig_1 & -ig_1 & -\Gamma_{\sigma_1} & -\frac{\gamma_{12}}{2} & 0 & 0 & 0 & -\frac{\gamma_{12}}{2}\\
  			0 & 0 & -ig_2 & -\frac{\gamma_{12}}{2} & i(\omega_{\sigma_1}\!\!-\!\!\omega_{\sigma_2}\!)\!\!-\!\!\frac{\Gamma_{\sigma_1}\!+\!\Gamma_2}{2} & ig_1 & 0 & -\frac{\gamma_{12}}{2} & 0\\
  			-ig_2 & -\frac{\gamma_{12}}{2} & 0 & 0 & ig_1 & i(\omega_a\!\!-\!\!\omega_{\sigma_2}\!)\!\!-\!\!\frac{\Gamma_a\!+\!\Gamma_{\sigma_2}}{2} & 0 & ig_2 & 0\\
  				ig_2 & 0 & -\frac{\gamma_{12}}{2} & 0 & 0 & 0 & i(\omega_{\sigma_2}\!\!-\!\!\omega_a\!)\!\!-\!\!\frac{\Gamma_{\sigma_1}\!+\!\Gamma_{\sigma_2}}{2} & -ig_2 & -ig_1\\
  				0&0&0&0& -\frac{\gamma_{12}}{2}&ig_2&-ig_2&-\Gamma_{\sigma_2}&-\frac{\gamma_{12}}{2}\\
  				0&ig_2&0&-\frac{\gamma_{12}}{2}&0&0&-ig_1&-\frac{\gamma_{12}}{2}&i(\omega_{\sigma_2}\!\!-\!\!\omega_{\sigma_1}\!)\!-\!\frac{\Gamma_{\sigma_1}\!+\!\Gamma_{\sigma_2}}{2}\\
  				\end{pmatrix}\notag\\
  				&\times\begin{pmatrix}
  				\langle a^\dagger(t) a(t) \rangle\\
  				\langle a^\dagger(t) \sigma_1(t) \rangle\\
  				\langle a(t) \sigma_1^\dagger(t) \rangle\\
  				\langle \sigma_1^\dagger(t) \sigma_1(t) \rangle\\
  				\langle \sigma_1^\dagger(t) \sigma_2(t) \rangle\\
  				\langle a^\dagger(t) \sigma_2(t)\rangle\\
  				\langle a(t) \sigma_2^\dagger(t) \rangle\\
  				\langle \sigma_2^\dagger(t) \sigma_2(t) \rangle\\
  				\langle \sigma_1(t) \sigma_2^\dagger(t) \rangle\\
  				\end{pmatrix},
  				\end{align}
                \end{widetext}
                Now, if the left-hand side of Eq. (\ref{alaki}) is equal to zero, the values of the element $\mathbf{u}_a(t,t)$ will be achieved.

\bibliography{bibliography}
\end{document}